\documentclass[prb, preprint,amsmath,amssymb]{revtex4-1}

\usepackage{amsmath}
\usepackage{graphics}
\usepackage{graphicx}
\usepackage{dcolumn}

\begin{document}

\title{Observation of the spin Seebeck effect in epitaxial Fe$_3$O$_4$
   thin films}
\author{R. Ramos$^{*,1}$\footnote[0]{$^{*)}$Electronic mail:
ramosr@unizar.es}, T. Kikkawa$^{2}$, K. Uchida$^{2,3}$, H. Adachi$^{4,5}$, I. Lucas$^{6}$, M.H. Aguirre$^{1}$, P. Algarabel$^{6}$, L. Morell\'{o}n$^{1,7}$, S. Maekawa$^{4,5}$, E. Saitoh$^{2,4,5,8}$ and M.R. Ibarra$^{1,7}$}
\affiliation{$^1$ Instituto de Nanociencia de Arag\'{o}n, Universidad de Zaragoza, E-50018 Zaragoza, Spain}
\affiliation{$^2$ Institute for Materials Research, Tohoku University, Sendai 980-8577, Japan}
\affiliation{$^3$ PRESTO, Japan Science and Technology Agency, Kawaguchi, Saitama 332-0012, Japan}
\affiliation{$^4$ Advanced Science Research Center, Japan Atomic Energy Agency, Tokai 319-1195, Japan}
\affiliation{$^5$ CREST, Japan Science and Technology Agency, Sanbancho, Tokyo 102-0075, Japan}
\affiliation{$^6$ Instituto de Ciencia de Materiales de Arag\'{o}n, Universidad de Zaragoza and Consejo Superior de Investigaciones Cient\'{i}ficas, 50009 Zaragoza, Spain}
\affiliation{$^7$ Departamento de F\'{i}sica de la Materia Condensada, Universidad de Zaragoza, E-50009 Zaragoza, Spain}
\affiliation{$^8$ WPI Advanced Institute for Materials Research, Tohoku University, Sendai 980-8577, Japan}

\date{\today}

\begin{abstract}
We report the first experimental observation of the spin Seebeck effect in magnetite thin films. The signal observed at temperatures above the Verwey transition is a contribution from both the anomalous Nernst (ANE) and spin Seebeck effects (SSE). The contribution from the ANE of the Fe$_3$O$_4$ layer to the SSE is found to be negligible due to the resistivity difference between Fe$_3$O$_4$ and Pt layers. Below the Verwey transition the SSE is free from the ANE of the ferromagnetic layer and it is also found to dominate over the ANE due to magnetic proximity effect on the Pt layer.
\\

\end{abstract}

\maketitle

\noindent

\begin{figure}[htp!] \center
    \includegraphics {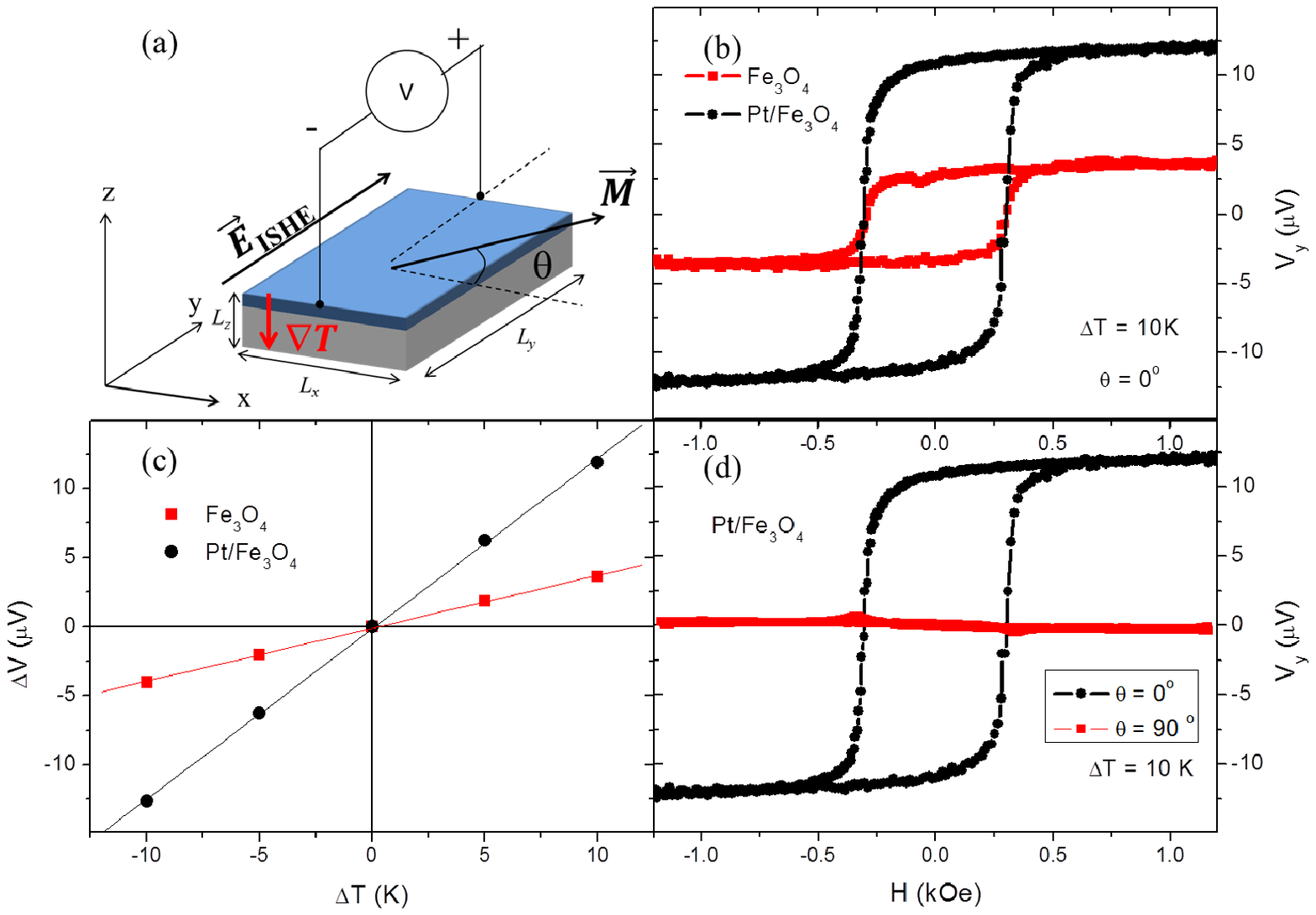}
  \caption{(Color online) (a) Schematic illustration of the measurement setup. (b) Obtained results for the Fe$_3$O$_4$ film with and without the Pt detecting layer at room temperature. (c) Dependence of $\Delta V$ measured at different magnitudes of the temperature difference ($\Delta T$) across the sample with and without Pt layer at room temperature. (d) Angular dependence of the measured voltage (V$_{y}$) in the Pt/Fe$_3$O$_4$ system at room temperature.}
  \label{fig1}
\end{figure}

\begin{figure}[htp!] \center
    \includegraphics {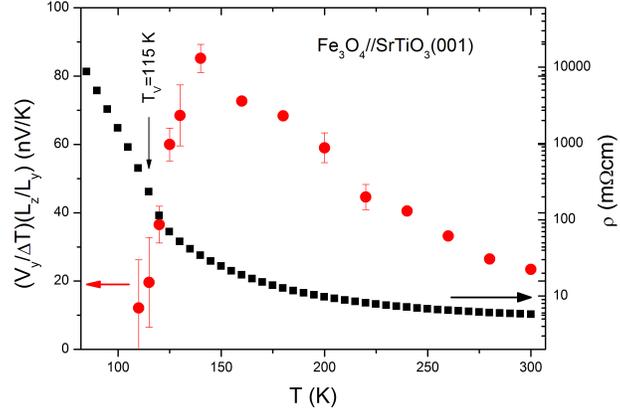}
  \caption{(Color online) Temperature dependence of the electrical resistivity (squares) and the geometrically corrected anomalous Nernst voltage  normalized with the applied temperature difference (circles) for the Fe$_3$O$_4$(001)//SrTiO$_3$(001) sample.}
  \label{fig2}
\end{figure}

\begin{figure} [htp!]
  \begin{center}
    \includegraphics {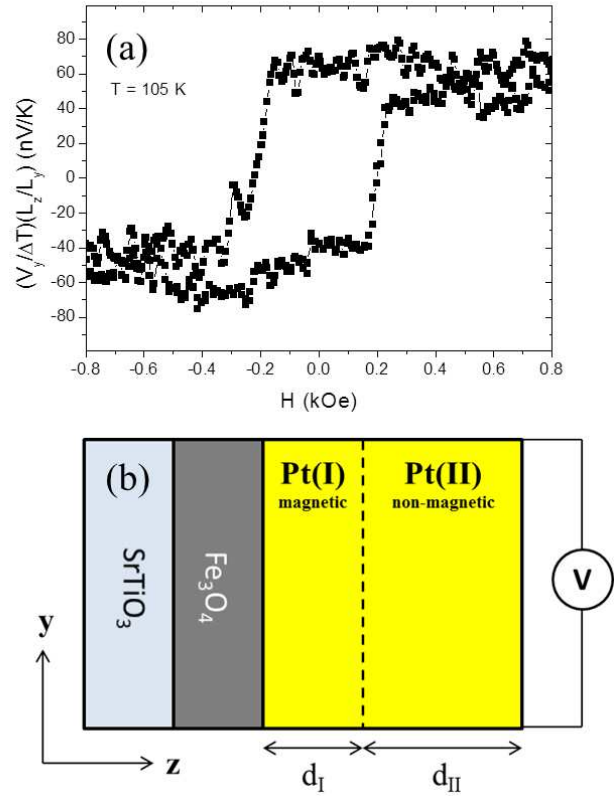}
  \end{center}
  \caption{(Color online) (a) Magnetic field dependence of the spin Seebeck effect measured at 105 K. (b) Schematic used to estimate the ANE due to magnetic proximity in the Pt layer.} \label{fig3:SSE}
\end{figure}

\newpage
Thermoelectric effects result from the combination between charge and heat current in suitable materials having applications as electric cooling systems or thermal power generators. Despite decades of research into themoelectric materials and properties, the efficiency of thermoelectric devices has remained low due to the interdependence of the Seebeck voltage, $S$, the resistivity, $\rho$, and the thermal conductivity, $\kappa$.\cite{Heremans2008, Jin2012} One promising approach to overcome this problem and increase the versatility of thermoelectric devices involves exploiting the spin of the electron, in addition to its charge and heat transport properties. This is the main interest of spin caloritronics.\cite{bauer:spinCaloritronics, maekawa:spinCaloritronics, Scharf2012, Hatami2009, LeBreton2011, Yu2010, Costache2012} Since the discovery of the spin Seebeck effect (SSE)\cite{uchida:nat2008} this field has been the focus of intensive theoretical and experimental research, with the recent detection among others of the spin-dependent Seebeck\cite{spin-depSeebeck} and Peltier effects.\cite{FlipseJ.2012} The SSE consists in the generation of a spin voltage in a ferromagnet as a result of an applied thermal gradient in magnetic materials, this spin voltage can be detected with an adjacent paramagnetic metal by means of the inverse spin Hall effect (ISHE).\cite{saitoh:apl-ISHE} Since its discovery in permalloy,\cite{uchida:nat2008} the SSE has been measured in spin polarized metals,\cite{Uchida2010} semiconductors,\cite{jaworsky:natmat2010, jaworsky:prl, Jaworski2012} and insulators.\cite{uchida:sse-insulator} In contrast to conventional thermoelectrics this effect offers the possibility of a new approach for all-solid state energy conversion devices, since it involves properties of at least two different materials that can be optimized independently. One example is the recent enhancement of the spin Seebeck effect by implementation of a spin Hall thermopile structure.\cite{Uchida:thermopile}\\
The SSE is explained in terms of a spin current injected from the ferromagnet (FM) into the paramagnetic metal (PM), which is scattered by the ISHE, generating an electric field $\vec{E}_{ISHE}$ given by
\begin{equation}
\label{eq:Eishe}
  \vec{E}_{ISHE}=\frac{\theta_{SH}\rho}{A}(\frac{2e}{\hbar})\vec{J_S} \times \vec{\sigma}
\end{equation}
where $\theta_{SH}$, $\rho$, $A$, $e$, $\vec{J_S}$ and  $\vec{\sigma}$ are the spin-Hall angle of PM, electric resistivity of PM, contact area between FM and PM, electron charge, spin current across the FM/PM interface and the spin polarization of FM respectively. The spin current $\vec{J_S}$ is generated as the result of a thermal non-equilibrium between magnons in FM and a conduction-electron spin accumulation in PM, which interact through the s-d exchange coupling $J_{sd}$ at the FM/PM interface. Using the linear-response approach, the spin current injected into PM is calculated to be\cite{Adachi2011, 2012arXiv_adachi}
\begin{equation}
\label{eq:SSE}
  J_s=-G_Sk_B(T^*_{FM}-T^*_{PM})
\end{equation}
 where $T^*_{FM}$ and $T^*_{PM}$ are the effective magnon temperature in FM and the effective conduction-electron temperature in PM, and $G_S = \frac{J^2_{sd}S_0\chi_N\tau_{sf}}{\hbar}$ with $S_0$, $\chi_N$ and $\tau_{sf}$ being respectively the size of the localized spin in FM, the paramagnetic susceptibility and spin flip relaxation time in PM. A similar interpretation has been developed using the scattering approach.\cite{Xiao2010}\\
 In this letter we report the first experimental observation of the spin Seebeck effect in magnetite. Magnetite is a ferrimagnetic oxide with a predicted half metallic character and a high Curie temperature (858 K), these properties have inspired investigations for possible spintronic applications,\cite{Sinova2012} therefore films and heterostructures of this material have been grown by several techniques.\cite{Ramos2008, Balakrishnan2004, Wu2010, Fernandez-Pacheco2009} Besides, magnetite possesses a metal-insulator transition at around 121 K, known as the Verwey transition.\cite{Walz2002}\\
A Fe$_3$O$_4$ (001) film (FM) of thickness $t_{I}$ = 50 nm was deposited on a SrTiO$_3$ (001) substrate of thickness $t_{STO}$ = 0.5 mm, by pulsed laser deposition (PLD) using a KrF excimer laser with 248 nm wavelength, 10 Hz repetition rate, and 3x10$^9$ W/cm$^2$ irradiance in an ultrahigh-vacuum chamber. The film thickness was measured by x-ray reflectivity (XRR). The Verwey transition temperature of the film was measured with SQUID and four probe resistivity measurements and has a value of $T_V$=115 K and the room temperature resistivity is 5 m$\Omega$ cm. Further details on the growth and characterization can be found elsewhere.\cite{Orna2010}\\
The SSE was measured using the so called longitudinal configuration\cite{Uchida:jap2012} (see Fig.\ref{fig1}(a)), a temperature gradient ($\nabla T$) is applied in the $\mp z$ direction, generating a temperature difference ($\pm \Delta T$) between the bottom and the top of the sample, with temperatures $T \pm \Delta T$ and $T$ respectively. The voltage ($V_{y}$) is measured along the $y$ direction, while a sweeping magnetic field is applied at an angle $\theta$ with respect to the $x$ direction. This configuration is normally used for insulating samples, for measurements performed on electrically conductive samples with this geometry, the anomalous Nernst effect (ANE)\cite{PRL_108_106602} is also measured along with the SSE. In order to separate the contribution of the ANE, simultaneous measurements of both effects were performed on the same Fe$_3$O$_4$ film, to do this, two equal pieces with a length of $L_y$=8 mm and a width of $L_x$=4 mm were cut and a Pt layer (PM) of thickness $t_{II}$ = 5 nm was deposited on one of them, both samples were loaded at the same time and kept under the same experimental conditions.\\
Figure \ref{fig1}(b) shows the results obtained at 300 K on the magnetic field dependence of the transversal voltage ($V_{y}$), measured on samples with and without the Pt layer. It is interesting to observe a strong enhancement of the signal upon placement of the spin detection layer (5 nm Pt), which is increased by $\sim$ 4 times with respect to that observed with no Pt layer. The fact that the resistivity at room temperature of the magnetite layer is 5 m$\Omega$cm which is about 2 orders of magnitude larger than the resistivity of the Pt layer, suggests that if there was no thermally induced spin pumping from the FM to the PM, the anomalous Nernst voltage on the Fe$_3$O$_4$ layer would be strongly suppressed by the Pt top layer. Therefore the voltage signal in the Pt/Fe$_3$O$_4$ sample must be dominated by the SSE. To estimate the suppression of the ANE signal upon placement of the Pt layer, we consider the expression from the electron transport theory $J^i_m=\sigma^{ij}_mE^j-\alpha^{ik}_m\nabla_kT$, where $J^i_m$ stands for the electron current, $E^j$ is the electric field and $\nabla_kT$ is the applied thermal gradient, the coefficients $\sigma^{ij}_m$ and $\alpha^{ik}_m$ are the elements of the conductivity and thermopower tensor respectively. Under our experimental conditions we obtain the following expressions:
\begin{equation}
\begin{split}
\label{eq:transport}
J^z_I=\sigma^{zz}_IE^z+\sigma^{zy}_IE^y-\alpha^{zz}_I(\nabla_zT)_I\\
J^y_I=\sigma^{yz}_IE^z+\sigma^{yy}_IE^y-\alpha^{zy}_I(\nabla_zT)_I\\
J^z_{II}=\sigma^{zz}_{II}E^z-\alpha^{zz}_{II}(\nabla_zT)_{II}\\
J^y_{II}=\sigma^{yy}_{II}E^y\\
\end{split}
\end{equation}
where $m=I$ and $m=II$ describe the FM (in the experiment, Fe$_3$O$_4$) and PM (in the experiment, Pt) respectively. Considering the open circuit condition: $I^z=A_0J^z_{I}=A_0J^z_{II}=0$ and $I^y=S_{I}J^y_{I}+S_{II}J^y_{II}=0$, with $A_0 = L_x L_y$ and $S_m = L_x t_m$ being the area of the film with normal to the $z$ and $y$ direction respectively. The following expresssion for the transversal  component $E^y$ is obtained:
\begin{equation}
\label{eq:ANE_short}
  E^y=(\frac{r}{1+r})E_{ANE}
\end{equation}
where $ r=\frac{\rho_{Pt}}{\rho_{Fe_3O_4}}\frac{t_{Fe_3O_4}}{t_{Pt}}$ and $E_{ANE}=[\frac{\alpha^{yz}}{\sigma^{zz}}-(\frac{\sigma^{yz}}{\sigma^{zz}}\frac{\alpha^{zz}}{\sigma^{zz}})](\nabla_z T)_{Fe_3O_4}$ is the anomalous Nernst signal measured in the Fe$_3$O$_4$ film. Considering a resistivity value of 4.8x10$^{-7}$ $\Omega$m for a Pt film grown under similar conditions\cite{Kirihara2012} and 5x10$^{-5}$ $\Omega$m for the 50 nm Fe$_3$O$_4$ film, we obtain that the ANE signal in the Pt/Fe$_3$O$_4$ sample is reduced to 10 \% of the ANE signal in the Fe$_3$O$_4$ sample, giving a contribution of about 3\% to the total observed signal $V_y = E^y L_y$ in Pt/Fe$_3$O$_4$. This is a strong indication that the spin Seebeck effect is the dominant contribution to the observed voltage. Measurements at different magnitudes of the thermal gradient with the magnetic field applied in the $x$ direction, were also performed, figure \ref{fig1}(c) shows the linear dependence of $\Delta V = (V_y(+H)-V_y(-H))/2)$ with $\Delta T$, as it is expected. From the slope of this curves we can extract the coefficients for the SSE and ANE, for the Pt/Fe$_3$O$_4$ and Fe$_3$O$_4$ sample, respectively. In order to evaluate the magnitude of the spin Seebeck effect we define the coefficient $S_{zy}=(\Delta V/\Delta T)(L_z/L_y)$, where $L_z$ is the length of the sample in the direction of the applied temperature gradient (see Fig. \ref{fig1}(a)). This coefficient is independent of the sample size and enables quantitative comparison between experiments with different geometries.\cite{Kikkawa2012} Considering the correction for the shorting effect of the ANE in Fe$_3$O$_4$ due to the Pt layer, described above, we obtain a value of the SSE at room temperature of $S_{zy}$ = 74 $\pm$ 1 nV/K. Measurements on the dependence of the direction of the applied magnetic field were also performed (see Fig. \ref{fig1}(d)), it can be observed that when the field is parallel to the direction in which the voltage is measured $\theta = 90 ^{\circ}$, the voltage vanishes, in agreement with Eq. (\ref{eq:Eishe}).\\
The longitudinal SSE configuration is normally used for insulating samples, therefore it is interesting to measure the effect at temperatures below the Verwey transition ($T_V$ = 115 K), after the sample undergoes the metal to insulator transition. Figure \ref{fig2} shows that the ANE in Fe$_3$O$_4$ could not be detected below $T$=110 K, due to the very high electrical resistance as a consequence of the strong reduction of charge carriers in the insulating phase. The temperature dependence of the ANE above the transition temperature resembles that of the thermal conductivity of the substrate,\cite{Yu2008} this will be the subject of subsequent studies. Here we will focus on the observation of the SSE below the Verwey transition. It is interesting to point out that despite the suppression of the ANE, the SSE can still be detected for $T < T_V$, as it is shown in Fig. \ref{fig3:SSE}(a) for $T$=105 K. This is a strong indication of the observation of pure spin Seebeck effect. We estimated a SSE coefficient ($S_{zy}$) at this temperature of about 52 $\pm$ 12 nV/K. It can be observed that there is a reduction of the SSE compared to the value observed at 300 K, this could be possibly related to the changes induced by the Verwey transition on the thermal\cite{2004:Salazar,Slack1962} and magnetic\cite{Fe3O4anis2012, 2006:McQueeney} properties of the film, which can affect the thermal spin pumping at the Fe$_3$O$_4$/Pt interface and therefore the observed SSE signal.\\
We proceed to estimate the contribution of the ANE due to magnetic proximity\cite{chien_magnproxPt} in the Pt layer. In order to do so, we consider the Pt layer to be divided into a magnetic and a non-magnetic region with thicknesses $d_I$ and $d_{II}$ respectively (see Fig. \ref{fig3:SSE}(b)). From the electron transport equations $J^i=\sigma^{ij}E^j-\alpha^{ij}\nabla_jT$, we obtain a similar expression to that observed in Eq. (\ref{eq:transport}), with $I$ and $II$ correspondent to the magnetic and non-magnetic regions of the Pt layer respectively. Considering the open circuit condition as described previously, we obtain the following expression for the transversal electric field:
\begin{equation}
\label{eq:ANE_MPE}
  E^y=(\frac{d_I}{d_I + d_{II}})E^*_{ANE}
\end{equation}
where
the parameter $\frac{d_I}{d_I + d_{II}}$ accounts for the shorting of the ANE of the magnetized region by the non-magnetized region within the Pt layer and $E^*_{ANE}=\{\frac{\alpha^{yz}}{\sigma^{zz}}-\frac{\sigma^{yz}}{\sigma^{zz}}\frac{\alpha^{zz}}{\sigma^{zz}}\}(\nabla T)_{Pt}=V^{MP}_{ANE}/L_y$ is the ANE of the magnetized Pt region. To estimate this quantity, we use the coefficients for the conductivity and thermopower tensor previously measured in FePt thin films\cite{ANE_FePt}. An estimation of the thermal gradient across the Pt thin film can be obtained by considering that the heat is conserved across the different interfaces of our system. For $t_{STO}\gg t_{Fe_3O_4}, t_{Pt}$ and using the tabulated values for thermal conductivity of SrTiO$_3$ and Pt,\cite{Lide2009} we obtained: $(\nabla T)_{Pt}\approx 3.75\times10^2\Delta T$ in K/m, with $\Delta T$ being the temperature difference generated between the hot and cold side of our sample. We obtained an expression for the upper limit of the contribution due to magnetic proximity in the Pt layer to the observed effect:
\begin{equation}
\label{eq:ANE_MPEn}
  \frac{V^{MP}_{ANE}}{\Delta T}\frac{L_z}{L_y} \leq 7.5 n [nV/K]
\end{equation}
where $n$ is the number of monolayers of Pt which are fully magnetized. If we consider one monolayer of Pt to be fully magnetized ($n$=1),\cite{magn_prox_NiPt} we obtain a contribution due to the proximity effect of 7.5 nV/K. This value is almost one order of magnitude smaller than the effect observed at 105 K and comparable to the error of the measurement, therefore the thermally induced voltage is clearly dominated by the SSE.\\
In conclusion we have observed for the first time the spin Seebeck effect in magnetite film. At room temperature, despite of the fact that the films are electrically conductive, the contribution of the ANE effect of magnetite only accounts for 3\% of the observed signal, due to the resistivity difference between Fe$_3$O$_4$ and Pt films. The ANE signal falls below the detection values below the Verwey transition, this points to a suppression of the ANE as a consequence of the reduction of charge carriers. Therefore at temperatures below T$_V$ the SSE signal is free from any contamination from ANE of the ferromagnetic layer. The effect of the magnetic proximity in Pt has also been evaluated below the metal-insulator transition and it has been observed to be one order of magnitude smaller than the measured signal, clearly showing that the SSE is the dominant contribution in our measurements.

\section{Acknowledgements}
R. Ramos and T. Kikkawa contributed equally on the thermal transport measurements. This work was supported by the Spanish Ministry of Science (trough projects I-2012-007, MAT2011-27553-C02, including FEDER funding) and the Arag\'{o}n Regional Government (project E26). This research was supported by Strategic International Cooperative Program, Japan Science and Technology Agency (JST), PRESTO-JST ``Phase Interfaces for Highly Efficient Energy Utilization'', CREST-JST ``Creation of Nanosystems with Novel Functions through Process Integration'', a Grant-in-Aid for Research Activity Start-up (24860003) from MEXT, Japan, a Grant-in-Aid for Scientific Research (A) (24244051) from MEXT, Japan, LC-IMR of Tohoku University, the Murata Science Foundation, and the Mazda Foundation.

\newpage

\newpage

\newpage

\begin{thebibliography}{42}
\expandafter\ifx\csname natexlab\endcsname\relax\def\natexlab#1{#1}\fi
\expandafter\ifx\csname bibnamefont\endcsname\relax
  \def\bibnamefont#1{#1}\fi
\expandafter\ifx\csname bibfnamefont\endcsname\relax
  \def\bibfnamefont#1{#1}\fi
\expandafter\ifx\csname citenamefont\endcsname\relax
  \def\citenamefont#1{#1}\fi
\expandafter\ifx\csname url\endcsname\relax
  \def\url#1{\texttt{#1}}\fi
\expandafter\ifx\csname urlprefix\endcsname\relax\def\urlprefix{URL }\fi
\providecommand{\bibinfo}[2]{#2}
\providecommand{\eprint}[2][]{\url{#2}}

\bibitem[{\citenamefont{Heremans et~al.}(2008)\citenamefont{Heremans, Jovovic,
  Toberer, Saramat, Kurosaki, Charoenphakdee, Yamanaka, and
  Snyder}}]{Heremans2008}
\bibinfo{author}{\bibfnamefont{J.~P.} \bibnamefont{Heremans}},
  \bibinfo{author}{\bibfnamefont{V.}~\bibnamefont{Jovovic}},
  \bibinfo{author}{\bibfnamefont{E.~S.} \bibnamefont{Toberer}},
  \bibinfo{author}{\bibfnamefont{A.}~\bibnamefont{Saramat}},
  \bibinfo{author}{\bibfnamefont{K.}~\bibnamefont{Kurosaki}},
  \bibinfo{author}{\bibfnamefont{A.}~\bibnamefont{Charoenphakdee}},
  \bibinfo{author}{\bibfnamefont{S.}~\bibnamefont{Yamanaka}}, \bibnamefont{and}
  \bibinfo{author}{\bibfnamefont{G.~J.} \bibnamefont{Snyder}},
  \bibinfo{journal}{Science} \textbf{\bibinfo{volume}{321}},
  \bibinfo{pages}{554} (\bibinfo{year}{2008}).

\bibitem[{\citenamefont{Jin et~al.}(2012)\citenamefont{Jin, Jaworski, and
  Heremans}}]{Jin2012}
\bibinfo{author}{\bibfnamefont{H.}~\bibnamefont{Jin}},
  \bibinfo{author}{\bibfnamefont{C.~M.} \bibnamefont{Jaworski}},
  \bibnamefont{and} \bibinfo{author}{\bibfnamefont{J.~P.}
  \bibnamefont{Heremans}}, \bibinfo{journal}{Appl. Phys. Lett.}
  \textbf{\bibinfo{volume}{101}}, \bibinfo{eid}{053904} (\bibinfo{year}{2012}).

\bibitem[{\citenamefont{Bauer et~al.}(2012)\citenamefont{Bauer, Saitoh, and van
  Wees}}]{bauer:spinCaloritronics}
\bibinfo{author}{\bibfnamefont{G.~E.~W.} \bibnamefont{Bauer}},
  \bibinfo{author}{\bibfnamefont{E.}~\bibnamefont{Saitoh}}, \bibnamefont{and}
  \bibinfo{author}{\bibfnamefont{B.~J.} \bibnamefont{van Wees}},
  \bibinfo{journal}{Nat. Mater.} \textbf{\bibinfo{volume}{11}},
  \bibinfo{pages}{391} (\bibinfo{year}{2012}).

\bibitem[{\citenamefont{Bauer et~al.}(2010)\citenamefont{Bauer, MacDonald, and
  Maekawa}}]{maekawa:spinCaloritronics}
\bibinfo{author}{\bibfnamefont{G.~E.~W.} \bibnamefont{Bauer}},
  \bibinfo{author}{\bibfnamefont{A.~H.} \bibnamefont{MacDonald}},
  \bibnamefont{and} \bibinfo{author}{\bibfnamefont{S.}~\bibnamefont{Maekawa}},
  \bibinfo{journal}{Solid State Commun.} \textbf{\bibinfo{volume}{150}},
  \bibinfo{pages}{459} (\bibinfo{year}{2010}).

\bibitem[{\citenamefont{Scharf et~al.}(2012)\citenamefont{Scharf,
  Matos-Abiague, \ifmmode \check{Z}\else \v{Z}\fi{}uti\ifmmode~\acute{c}\else
  \'{c}\fi{}, and Fabian}}]{Scharf2012}
\bibinfo{author}{\bibfnamefont{B.}~\bibnamefont{Scharf}},
  \bibinfo{author}{\bibfnamefont{A.}~\bibnamefont{Matos-Abiague}},
  \bibinfo{author}{\bibfnamefont{I.}~\bibnamefont{\ifmmode \check{Z}\else
  \v{Z}\fi{}uti\ifmmode~\acute{c}\else \'{c}\fi{}}}, \bibnamefont{and}
  \bibinfo{author}{\bibfnamefont{J.}~\bibnamefont{Fabian}},
  \bibinfo{journal}{Phys. Rev. B} \textbf{\bibinfo{volume}{85}},
  \bibinfo{pages}{085208} (\bibinfo{year}{2012}).

\bibitem[{\citenamefont{Hatami et~al.}(2009)\citenamefont{Hatami, Bauer, Zhang,
  and Kelly}}]{Hatami2009}
\bibinfo{author}{\bibfnamefont{M.}~\bibnamefont{Hatami}},
  \bibinfo{author}{\bibfnamefont{G.~E.~W.} \bibnamefont{Bauer}},
  \bibinfo{author}{\bibfnamefont{Q.}~\bibnamefont{Zhang}}, \bibnamefont{and}
  \bibinfo{author}{\bibfnamefont{P.~J.} \bibnamefont{Kelly}},
  \bibinfo{journal}{Phys. Rev. B} \textbf{\bibinfo{volume}{79}},
  \bibinfo{pages}{174426} (\bibinfo{year}{2009}).

\bibitem[{\citenamefont{Le~Breton et~al.}(2011)\citenamefont{Le~Breton, Sharma,
  Saito, Yuasa, and Jansen}}]{LeBreton2011}
\bibinfo{author}{\bibfnamefont{J.-C.} \bibnamefont{Le~Breton}},
  \bibinfo{author}{\bibfnamefont{S.}~\bibnamefont{Sharma}},
  \bibinfo{author}{\bibfnamefont{H.}~\bibnamefont{Saito}},
  \bibinfo{author}{\bibfnamefont{S.}~\bibnamefont{Yuasa}}, \bibnamefont{and}
  \bibinfo{author}{\bibfnamefont{R.}~\bibnamefont{Jansen}},
  \bibinfo{journal}{Nature} \textbf{\bibinfo{volume}{475}}, \bibinfo{pages}{82}
  (\bibinfo{year}{2011}).

\bibitem[{\citenamefont{Yu et~al.}(2010)\citenamefont{Yu, Granville, Yu, and
  Ansermet}}]{Yu2010}
\bibinfo{author}{\bibfnamefont{H.}~\bibnamefont{Yu}},
  \bibinfo{author}{\bibfnamefont{S.}~\bibnamefont{Granville}},
  \bibinfo{author}{\bibfnamefont{D.~P.} \bibnamefont{Yu}}, \bibnamefont{and}
  \bibinfo{author}{\bibfnamefont{J.-P.} \bibnamefont{Ansermet}},
  \bibinfo{journal}{Phys. Rev. Lett.} \textbf{\bibinfo{volume}{104}},
  \bibinfo{pages}{146601} (\bibinfo{year}{2010}).

\bibitem[{\citenamefont{Costache et~al.}(2012)\citenamefont{Costache, Bridoux,
  Neumann, and Valenzuela}}]{Costache2012}
\bibinfo{author}{\bibfnamefont{M.~V.} \bibnamefont{Costache}},
  \bibinfo{author}{\bibfnamefont{G.}~\bibnamefont{Bridoux}},
  \bibinfo{author}{\bibfnamefont{I.}~\bibnamefont{Neumann}}, \bibnamefont{and}
  \bibinfo{author}{\bibfnamefont{S.~O.} \bibnamefont{Valenzuela}},
  \bibinfo{journal}{Nat. Mater.} \textbf{\bibinfo{volume}{11}},
  \bibinfo{pages}{199} (\bibinfo{year}{2012}).

\bibitem[{\citenamefont{Uchida et~al.}({2008})\citenamefont{Uchida, Takahashi,
  Harii, Ieda, Koshibae, Ando, Maekawa, and Saitoh}}]{uchida:nat2008}
\bibinfo{author}{\bibfnamefont{K.}~\bibnamefont{Uchida}},
  \bibinfo{author}{\bibfnamefont{S.}~\bibnamefont{Takahashi}},
  \bibinfo{author}{\bibfnamefont{K.}~\bibnamefont{Harii}},
  \bibinfo{author}{\bibfnamefont{J.}~\bibnamefont{Ieda}},
  \bibinfo{author}{\bibfnamefont{W.}~\bibnamefont{Koshibae}},
  \bibinfo{author}{\bibfnamefont{K.}~\bibnamefont{Ando}},
  \bibinfo{author}{\bibfnamefont{S.}~\bibnamefont{Maekawa}}, \bibnamefont{and}
  \bibinfo{author}{\bibfnamefont{E.}~\bibnamefont{Saitoh}},
  \bibinfo{journal}{Nature} \textbf{\bibinfo{volume}{{455}}},
  \bibinfo{pages}{{778}} (\bibinfo{year}{{2008}}).

\bibitem[{\citenamefont{Slachter et~al.}(2010)\citenamefont{Slachter, Bakker,
  Adam, and van Wees}}]{spin-depSeebeck}
\bibinfo{author}{\bibfnamefont{A.}~\bibnamefont{Slachter}},
  \bibinfo{author}{\bibfnamefont{F.~L.} \bibnamefont{Bakker}},
  \bibinfo{author}{\bibfnamefont{J.-P.} \bibnamefont{Adam}}, \bibnamefont{and}
  \bibinfo{author}{\bibfnamefont{B.~J.} \bibnamefont{van Wees}},
  \bibinfo{journal}{Nat. Phys.} \textbf{\bibinfo{volume}{6}},
  \bibinfo{pages}{879} (\bibinfo{year}{2010}).

\bibitem[{\citenamefont{Flipse et~al.}(2012)\citenamefont{Flipse, Bakker,
  Slachter, Dejene, and van Wees}}]{FlipseJ.2012}
\bibinfo{author}{\bibfnamefont{J.}~\bibnamefont{Flipse}},
  \bibinfo{author}{\bibfnamefont{F.~L.} \bibnamefont{Bakker}},
  \bibinfo{author}{\bibfnamefont{A.}~\bibnamefont{Slachter}},
  \bibinfo{author}{\bibfnamefont{F.~K.} \bibnamefont{Dejene}},
  \bibnamefont{and} \bibinfo{author}{\bibfnamefont{B.~J.} \bibnamefont{van
  Wees}}, \bibinfo{journal}{Nat. Nano.} \textbf{\bibinfo{volume}{7}},
  \bibinfo{pages}{166} (\bibinfo{year}{2012}).

\bibitem[{\citenamefont{Saitoh et~al.}(2006)\citenamefont{Saitoh, Ueda,
  Miyajima, and Tatara}}]{saitoh:apl-ISHE}
\bibinfo{author}{\bibfnamefont{E.}~\bibnamefont{Saitoh}},
  \bibinfo{author}{\bibfnamefont{M.}~\bibnamefont{Ueda}},
  \bibinfo{author}{\bibfnamefont{H.}~\bibnamefont{Miyajima}}, \bibnamefont{and}
  \bibinfo{author}{\bibfnamefont{G.}~\bibnamefont{Tatara}},
  \bibinfo{journal}{Appl. Phys. Lett.} \textbf{\bibinfo{volume}{88}},
  \bibinfo{eid}{182509} (\bibinfo{year}{2006}).

\bibitem[{\citenamefont{Uchida et~al.}(2010{\natexlab{a}})\citenamefont{Uchida,
  Ota, Harii, Ando, Nakayama, and Saitoh}}]{Uchida2010}
\bibinfo{author}{\bibfnamefont{K.}~\bibnamefont{Uchida}},
  \bibinfo{author}{\bibfnamefont{T.}~\bibnamefont{Ota}},
  \bibinfo{author}{\bibfnamefont{K.}~\bibnamefont{Harii}},
  \bibinfo{author}{\bibfnamefont{K.}~\bibnamefont{Ando}},
  \bibinfo{author}{\bibfnamefont{H.}~\bibnamefont{Nakayama}}, \bibnamefont{and}
  \bibinfo{author}{\bibfnamefont{E.}~\bibnamefont{Saitoh}},
  \bibinfo{journal}{J. of Appl. Phys.} \textbf{\bibinfo{volume}{107}},
  \bibinfo{eid}{09A951} (\bibinfo{year}{2010}{\natexlab{a}}).

\bibitem[{\citenamefont{Jaworski et~al.}(2010)\citenamefont{Jaworski, Yang,
  Mack, Awschalom, Heremans, and Myers}}]{jaworsky:natmat2010}
\bibinfo{author}{\bibfnamefont{C.~M.} \bibnamefont{Jaworski}},
  \bibinfo{author}{\bibfnamefont{J.}~\bibnamefont{Yang}},
  \bibinfo{author}{\bibfnamefont{S.}~\bibnamefont{Mack}},
  \bibinfo{author}{\bibfnamefont{D.~D.} \bibnamefont{Awschalom}},
  \bibinfo{author}{\bibfnamefont{J.~P.} \bibnamefont{Heremans}},
  \bibnamefont{and} \bibinfo{author}{\bibfnamefont{R.~C.} \bibnamefont{Myers}},
  \bibinfo{journal}{Nat. Mater.} \textbf{\bibinfo{volume}{9}},
  \bibinfo{pages}{898} (\bibinfo{year}{2010}).

\bibitem[{\citenamefont{Jaworski et~al.}(2011)\citenamefont{Jaworski, Yang,
  Mack, Awschalom, Myers, and Heremans}}]{jaworsky:prl}
\bibinfo{author}{\bibfnamefont{C.~M.} \bibnamefont{Jaworski}},
  \bibinfo{author}{\bibfnamefont{J.}~\bibnamefont{Yang}},
  \bibinfo{author}{\bibfnamefont{S.}~\bibnamefont{Mack}},
  \bibinfo{author}{\bibfnamefont{D.~D.} \bibnamefont{Awschalom}},
  \bibinfo{author}{\bibfnamefont{R.~C.} \bibnamefont{Myers}}, \bibnamefont{and}
  \bibinfo{author}{\bibfnamefont{J.~P.} \bibnamefont{Heremans}},
  \bibinfo{journal}{Phys. Rev. Lett.} \textbf{\bibinfo{volume}{106}},
  \bibinfo{pages}{186601} (\bibinfo{year}{2011}).

\bibitem[{\citenamefont{Jaworski et~al.}(2012)\citenamefont{Jaworski, Myers,
  Johnston-Halperin, and Heremans}}]{Jaworski2012}
\bibinfo{author}{\bibfnamefont{C.~M.} \bibnamefont{Jaworski}},
  \bibinfo{author}{\bibfnamefont{R.~C.} \bibnamefont{Myers}},
  \bibinfo{author}{\bibfnamefont{E.}~\bibnamefont{Johnston-Halperin}},
  \bibnamefont{and} \bibinfo{author}{\bibfnamefont{J.~P.}
  \bibnamefont{Heremans}}, \bibinfo{journal}{Nature}
  \textbf{\bibinfo{volume}{487}}, \bibinfo{pages}{210} (\bibinfo{year}{2012}).

\bibitem[{\citenamefont{Uchida et~al.}(2010{\natexlab{b}})\citenamefont{Uchida,
  Xiao, Adachi, Ohe, Takahashi, Ieda, Ota, Kajiwara, Umezawa, Kawai
  et~al.}}]{uchida:sse-insulator}
\bibinfo{author}{\bibfnamefont{K.}~\bibnamefont{Uchida}},
  \bibinfo{author}{\bibfnamefont{J.}~\bibnamefont{Xiao}},
  \bibinfo{author}{\bibfnamefont{H.}~\bibnamefont{Adachi}},
  \bibinfo{author}{\bibfnamefont{J.}~\bibnamefont{Ohe}},
  \bibinfo{author}{\bibfnamefont{S.}~\bibnamefont{Takahashi}},
  \bibinfo{author}{\bibfnamefont{J.}~\bibnamefont{Ieda}},
  \bibinfo{author}{\bibfnamefont{T.}~\bibnamefont{Ota}},
  \bibinfo{author}{\bibfnamefont{Y.}~\bibnamefont{Kajiwara}},
  \bibinfo{author}{\bibfnamefont{H.}~\bibnamefont{Umezawa}},
  \bibinfo{author}{\bibfnamefont{H.}~\bibnamefont{Kawai}},
  \bibnamefont{et~al.}, \bibinfo{journal}{Nat. Mater.}
  \textbf{\bibinfo{volume}{9}}, \bibinfo{pages}{894}
  (\bibinfo{year}{2010}{\natexlab{b}}).

\bibitem[{\citenamefont{Uchida et~al.}(2012{\natexlab{a}})\citenamefont{Uchida,
  Nonaka, Yoshino, Kikkawa, Kikuchi, and Saitoh}}]{Uchida:thermopile}
\bibinfo{author}{\bibfnamefont{K.}~\bibnamefont{Uchida}},
  \bibinfo{author}{\bibfnamefont{T.}~\bibnamefont{Nonaka}},
  \bibinfo{author}{\bibfnamefont{T.}~\bibnamefont{Yoshino}},
  \bibinfo{author}{\bibfnamefont{T.}~\bibnamefont{Kikkawa}},
  \bibinfo{author}{\bibfnamefont{D.}~\bibnamefont{Kikuchi}}, \bibnamefont{and}
  \bibinfo{author}{\bibfnamefont{E.}~\bibnamefont{Saitoh}},
  \bibinfo{journal}{Appl. Phys. Express} \textbf{\bibinfo{volume}{5}},
  \bibinfo{pages}{093001} (\bibinfo{year}{2012}{\natexlab{a}}).

\bibitem[{\citenamefont{Adachi et~al.}(2011)\citenamefont{Adachi, Ohe,
  Takahashi, and Maekawa}}]{Adachi2011}
\bibinfo{author}{\bibfnamefont{H.}~\bibnamefont{Adachi}},
  \bibinfo{author}{\bibfnamefont{J.-i.} \bibnamefont{Ohe}},
  \bibinfo{author}{\bibfnamefont{S.}~\bibnamefont{Takahashi}},
  \bibnamefont{and} \bibinfo{author}{\bibfnamefont{S.}~\bibnamefont{Maekawa}},
  \bibinfo{journal}{Phys. Rev. B} \textbf{\bibinfo{volume}{83}},
  \bibinfo{pages}{094410} (\bibinfo{year}{2011}).

\bibitem[{\citenamefont{{Adachi} et~al.}(2012)\citenamefont{{Adachi}, {Uchida},
  {Saitoh}, and {Maekawa}}}]{2012arXiv_adachi}
\bibinfo{author}{\bibfnamefont{H.}~\bibnamefont{{Adachi}}},
  \bibinfo{author}{\bibfnamefont{K.}~\bibnamefont{{Uchida}}},
  \bibinfo{author}{\bibfnamefont{E.}~\bibnamefont{{Saitoh}}}, \bibnamefont{and}
  \bibinfo{author}{\bibfnamefont{S.}~\bibnamefont{{Maekawa}}},
  \bibinfo{journal}{ArXiv e-prints}  (\bibinfo{year}{2012}),
  \eprint{1209.6407}.

\bibitem[{\citenamefont{Xiao et~al.}(2010)\citenamefont{Xiao, Bauer, Uchida,
  Saitoh, and Maekawa}}]{Xiao2010}
\bibinfo{author}{\bibfnamefont{J.}~\bibnamefont{Xiao}},
  \bibinfo{author}{\bibfnamefont{G.~E.~W.} \bibnamefont{Bauer}},
  \bibinfo{author}{\bibfnamefont{K.}~\bibnamefont{Uchida}},
  \bibinfo{author}{\bibfnamefont{E.}~\bibnamefont{Saitoh}}, \bibnamefont{and}
  \bibinfo{author}{\bibfnamefont{S.}~\bibnamefont{Maekawa}},
  \bibinfo{journal}{Phys. Rev. B} \textbf{\bibinfo{volume}{81}},
  \bibinfo{pages}{214418} (\bibinfo{year}{2010}).

\bibitem[{\citenamefont{Sinova and Zutic}(2012)}]{Sinova2012}
\bibinfo{author}{\bibfnamefont{J.}~\bibnamefont{Sinova}} \bibnamefont{and}
  \bibinfo{author}{\bibfnamefont{I.}~\bibnamefont{Zutic}},
  \bibinfo{journal}{Nat. Mater.} \textbf{\bibinfo{volume}{11}},
  \bibinfo{pages}{368} (\bibinfo{year}{2012}).

\bibitem[{\citenamefont{Ramos et~al.}(2008)\citenamefont{Ramos, Arora, and
  Shvets}}]{Ramos2008}
\bibinfo{author}{\bibfnamefont{R.}~\bibnamefont{Ramos}},
  \bibinfo{author}{\bibfnamefont{S.~K.} \bibnamefont{Arora}}, \bibnamefont{and}
  \bibinfo{author}{\bibfnamefont{I.~V.} \bibnamefont{Shvets}},
  \bibinfo{journal}{Phys. Rev. B} \textbf{\bibinfo{volume}{78}},
  \bibinfo{pages}{214402} (\bibinfo{year}{2008}).

\bibitem[{\citenamefont{Balakrishnan et~al.}(2004)\citenamefont{Balakrishnan,
  Arora, and Shvets}}]{Balakrishnan2004}
\bibinfo{author}{\bibfnamefont{K.}~\bibnamefont{Balakrishnan}},
  \bibinfo{author}{\bibfnamefont{S.~K.} \bibnamefont{Arora}}, \bibnamefont{and}
  \bibinfo{author}{\bibfnamefont{I.~V.} \bibnamefont{Shvets}},
  \bibinfo{journal}{J. Phys.: Condens. Matter.} \textbf{\bibinfo{volume}{16}},
  \bibinfo{pages}{5387} (\bibinfo{year}{2004}).

\bibitem[{\citenamefont{Wu et~al.}(2010)\citenamefont{Wu, Abid, Chun, Ramos,
  Mryasov, and Shvets}}]{Wu2010}
\bibinfo{author}{\bibfnamefont{H.-C.} \bibnamefont{Wu}},
  \bibinfo{author}{\bibfnamefont{M.}~\bibnamefont{Abid}},
  \bibinfo{author}{\bibfnamefont{B.~S.} \bibnamefont{Chun}},
  \bibinfo{author}{\bibfnamefont{R.}~\bibnamefont{Ramos}},
  \bibinfo{author}{\bibfnamefont{O.~N.} \bibnamefont{Mryasov}},
  \bibnamefont{and} \bibinfo{author}{\bibfnamefont{I.~V.}
  \bibnamefont{Shvets}}, \bibinfo{journal}{Nano Lett.}
  \textbf{\bibinfo{volume}{10}}, \bibinfo{pages}{1132} (\bibinfo{year}{2010}).

\bibitem[{\citenamefont{Fernandez-Pacheco
  et~al.}(2009)\citenamefont{Fernandez-Pacheco, Orna, Teresa, Algarabel,
  Morell\'{o}n, Pardo, Ibarra, Kampert, and Zeitler}}]{Fernandez-Pacheco2009}
\bibinfo{author}{\bibfnamefont{A.}~\bibnamefont{Fernandez-Pacheco}},
  \bibinfo{author}{\bibfnamefont{J.}~\bibnamefont{Orna}},
  \bibinfo{author}{\bibfnamefont{J.~M.~D.} \bibnamefont{Teresa}},
  \bibinfo{author}{\bibfnamefont{P.~A.} \bibnamefont{Algarabel}},
  \bibinfo{author}{\bibfnamefont{L.}~\bibnamefont{Morell\'{o}n}},
  \bibinfo{author}{\bibfnamefont{J.~A.} \bibnamefont{Pardo}},
  \bibinfo{author}{\bibfnamefont{M.~R.} \bibnamefont{Ibarra}},
  \bibinfo{author}{\bibfnamefont{E.}~\bibnamefont{Kampert}}, \bibnamefont{and}
  \bibinfo{author}{\bibfnamefont{U.}~\bibnamefont{Zeitler}},
  \bibinfo{journal}{Appl. Phys. Lett.} \textbf{\bibinfo{volume}{95}},
  \bibinfo{eid}{262108} (\bibinfo{year}{2009}).

\bibitem[{\citenamefont{Walz}(2002)}]{Walz2002}
\bibinfo{author}{\bibfnamefont{F.}~\bibnamefont{Walz}}, \bibinfo{journal}{J.
  Phys.: Condens. Matter.} \textbf{\bibinfo{volume}{14}}, \bibinfo{pages}{R285}
  (\bibinfo{year}{2002}).

\bibitem[{\citenamefont{Orna et~al.}(2010)\citenamefont{Orna, Algarabel,
  Morell{\'o}n, Pardo, de~Teresa, L{\'o}pez~Ant{\'o}n, Bartolom{\'e},
  Garc{\'i}a, Bartolom{\'e}, Cezar et~al.}}]{Orna2010}
\bibinfo{author}{\bibfnamefont{J.}~\bibnamefont{Orna}},
  \bibinfo{author}{\bibfnamefont{P.~A.} \bibnamefont{Algarabel}},
  \bibinfo{author}{\bibfnamefont{L.}~\bibnamefont{Morell{\'o}n}},
  \bibinfo{author}{\bibfnamefont{J.~A.} \bibnamefont{Pardo}},
  \bibinfo{author}{\bibfnamefont{J.~M.} \bibnamefont{de~Teresa}},
  \bibinfo{author}{\bibfnamefont{R.}~\bibnamefont{L{\'o}pez~Ant{\'o}n}},
  \bibinfo{author}{\bibfnamefont{F.}~\bibnamefont{Bartolom{\'e}}},
  \bibinfo{author}{\bibfnamefont{L.~M.} \bibnamefont{Garc{\'i}a}},
  \bibinfo{author}{\bibfnamefont{J.}~\bibnamefont{Bartolom{\'e}}},
  \bibinfo{author}{\bibfnamefont{J.~C.} \bibnamefont{Cezar}},
  \bibnamefont{et~al.}, \bibinfo{journal}{Phys. Rev. B}
  \textbf{\bibinfo{volume}{81}}, \bibinfo{pages}{144420}
  (\bibinfo{year}{2010}).

\bibitem[{\citenamefont{Uchida et~al.}(2012{\natexlab{b}})\citenamefont{Uchida,
  Ota, Adachi, Xiao, Nonaka, Kajiwara, Bauer, Maekawa, and
  Saitoh}}]{Uchida:jap2012}
\bibinfo{author}{\bibfnamefont{K.}~\bibnamefont{Uchida}},
  \bibinfo{author}{\bibfnamefont{T.}~\bibnamefont{Ota}},
  \bibinfo{author}{\bibfnamefont{H.}~\bibnamefont{Adachi}},
  \bibinfo{author}{\bibfnamefont{J.}~\bibnamefont{Xiao}},
  \bibinfo{author}{\bibfnamefont{T.}~\bibnamefont{Nonaka}},
  \bibinfo{author}{\bibfnamefont{Y.}~\bibnamefont{Kajiwara}},
  \bibinfo{author}{\bibfnamefont{G.~E.~W.} \bibnamefont{Bauer}},
  \bibinfo{author}{\bibfnamefont{S.}~\bibnamefont{Maekawa}}, \bibnamefont{and}
  \bibinfo{author}{\bibfnamefont{E.}~\bibnamefont{Saitoh}},
  \bibinfo{journal}{J. of Appl. Phys.} \textbf{\bibinfo{volume}{111}},
  \bibinfo{eid}{103903} (\bibinfo{year}{2012}{\natexlab{b}}).

\bibitem[{\citenamefont{Weiler et~al.}(2012)\citenamefont{Weiler, Althammer,
  Czeschka, Huebl, Wagner, Opel, Imort, Reiss, Thomas, Gross
  et~al.}}]{PRL_108_106602}
\bibinfo{author}{\bibfnamefont{M.}~\bibnamefont{Weiler}},
  \bibinfo{author}{\bibfnamefont{M.}~\bibnamefont{Althammer}},
  \bibinfo{author}{\bibfnamefont{F.~D.} \bibnamefont{Czeschka}},
  \bibinfo{author}{\bibfnamefont{H.}~\bibnamefont{Huebl}},
  \bibinfo{author}{\bibfnamefont{M.~S.} \bibnamefont{Wagner}},
  \bibinfo{author}{\bibfnamefont{M.}~\bibnamefont{Opel}},
  \bibinfo{author}{\bibfnamefont{I.-M.} \bibnamefont{Imort}},
  \bibinfo{author}{\bibfnamefont{G.}~\bibnamefont{Reiss}},
  \bibinfo{author}{\bibfnamefont{A.}~\bibnamefont{Thomas}},
  \bibinfo{author}{\bibfnamefont{R.}~\bibnamefont{Gross}},
  \bibnamefont{et~al.}, \bibinfo{journal}{Phys. Rev. Lett.}
  \textbf{\bibinfo{volume}{108}}, \bibinfo{pages}{106602}
  (\bibinfo{year}{2012}).

\bibitem[{\citenamefont{Kirihara et~al.}(2012)\citenamefont{Kirihara, Uchida,
  Kajiwara, Ishida, Nakamura, Manako, Saitoh, and Yorozu}}]{Kirihara2012}
\bibinfo{author}{\bibfnamefont{A.}~\bibnamefont{Kirihara}},
  \bibinfo{author}{\bibfnamefont{K.}~\bibnamefont{Uchida}},
  \bibinfo{author}{\bibfnamefont{Y.}~\bibnamefont{Kajiwara}},
  \bibinfo{author}{\bibfnamefont{M.}~\bibnamefont{Ishida}},
  \bibinfo{author}{\bibfnamefont{Y.}~\bibnamefont{Nakamura}},
  \bibinfo{author}{\bibfnamefont{T.}~\bibnamefont{Manako}},
  \bibinfo{author}{\bibfnamefont{E.}~\bibnamefont{Saitoh}}, \bibnamefont{and}
  \bibinfo{author}{\bibfnamefont{S.}~\bibnamefont{Yorozu}},
  \bibinfo{journal}{Nat. Mater.} \textbf{\bibinfo{volume}{11}},
  \bibinfo{pages}{686} (\bibinfo{year}{2012}).

\bibitem[{\citenamefont{{Kikkawa} et~al.}(2012)\citenamefont{{Kikkawa},
  {Uchida}, {Shiomi}, {Qiu}, {Hou}, {Tian}, {Nakayama}, {Jin}, and
  {Saitoh}}}]{Kikkawa2012}
\bibinfo{author}{\bibfnamefont{T.}~\bibnamefont{{Kikkawa}}},
  \bibinfo{author}{\bibfnamefont{K.}~\bibnamefont{{Uchida}}},
  \bibinfo{author}{\bibfnamefont{Y.}~\bibnamefont{{Shiomi}}},
  \bibinfo{author}{\bibfnamefont{Z.}~\bibnamefont{{Qiu}}},
  \bibinfo{author}{\bibfnamefont{D.}~\bibnamefont{{Hou}}},
  \bibinfo{author}{\bibfnamefont{D.}~\bibnamefont{{Tian}}},
  \bibinfo{author}{\bibfnamefont{H.}~\bibnamefont{{Nakayama}}},
  \bibinfo{author}{\bibfnamefont{X.-F.} \bibnamefont{{Jin}}}, \bibnamefont{and}
  \bibinfo{author}{\bibfnamefont{E.}~\bibnamefont{{Saitoh}}},
  \bibinfo{journal}{ArXiv e-prints}  (\bibinfo{year}{2012}),
  \eprint{1211.0139}.

\bibitem[{\citenamefont{Yu et~al.}(2008)\citenamefont{Yu, Scullin, Huijben,
  Ramesh, and Majumdar}}]{Yu2008}
\bibinfo{author}{\bibfnamefont{C.}~\bibnamefont{Yu}},
  \bibinfo{author}{\bibfnamefont{M.~L.} \bibnamefont{Scullin}},
  \bibinfo{author}{\bibfnamefont{M.}~\bibnamefont{Huijben}},
  \bibinfo{author}{\bibfnamefont{R.}~\bibnamefont{Ramesh}}, \bibnamefont{and}
  \bibinfo{author}{\bibfnamefont{A.}~\bibnamefont{Majumdar}},
  \bibinfo{journal}{Appl. Phys. Lett.} \textbf{\bibinfo{volume}{92}},
  \bibinfo{eid}{191911} (\bibinfo{year}{2008}).

\bibitem[{\citenamefont{Salazar et~al.}(2004)\citenamefont{Salazar, Oleaga,
  Wiechec, Tarnawski, and Kozlowski}}]{2004:Salazar}
\bibinfo{author}{\bibfnamefont{A.}~\bibnamefont{Salazar}},
  \bibinfo{author}{\bibfnamefont{A.}~\bibnamefont{Oleaga}},
  \bibinfo{author}{\bibfnamefont{A.}~\bibnamefont{Wiechec}},
  \bibinfo{author}{\bibfnamefont{Z.}~\bibnamefont{Tarnawski}},
  \bibnamefont{and}
  \bibinfo{author}{\bibfnamefont{A.}~\bibnamefont{Kozlowski}},
  \bibinfo{journal}{IEEE Trans. Magn.} \textbf{\bibinfo{volume}{40}},
  \bibinfo{pages}{2820 } (\bibinfo{year}{2004}).

\bibitem[{\citenamefont{Slack}(1962)}]{Slack1962}
\bibinfo{author}{\bibfnamefont{G.~A.} \bibnamefont{Slack}},
  \bibinfo{journal}{Phys. Rev.} \textbf{\bibinfo{volume}{126}},
  \bibinfo{pages}{427} (\bibinfo{year}{1962}).

\bibitem[{\citenamefont{\u{R}ezn\'{i}\u{c}ek
  et~al.}(2012)\citenamefont{\u{R}ezn\'{i}\u{c}ek, Chlan,
  \u{S}t\u{e}p\'{a}nkov\'{a}, Nov\'{a}k, and Mary\u{s}ko}}]{Fe3O4anis2012}
\bibinfo{author}{\bibfnamefont{R.}~\bibnamefont{\u{R}ezn\'{i}\u{c}ek}},
  \bibinfo{author}{\bibfnamefont{V.}~\bibnamefont{Chlan}},
  \bibinfo{author}{\bibfnamefont{H.}~\bibnamefont{\u{S}t\u{e}p\'{a}nkov\'{a}}},
  \bibinfo{author}{\bibfnamefont{P.}~\bibnamefont{Nov\'{a}k}},
  \bibnamefont{and}
  \bibinfo{author}{\bibfnamefont{M.}~\bibnamefont{Mary\u{s}ko}},
  \bibinfo{journal}{J. Phys.: Condens. Matter} \textbf{\bibinfo{volume}{24}},
  \bibinfo{pages}{055501} (\bibinfo{year}{2012}).

\bibitem[{\citenamefont{McQueeney et~al.}(2006)\citenamefont{McQueeney,
  Yethiraj, Montfrooij, Gardner, Metcalf, and Honig}}]{2006:McQueeney}
\bibinfo{author}{\bibfnamefont{R.~J.} \bibnamefont{McQueeney}},
  \bibinfo{author}{\bibfnamefont{M.}~\bibnamefont{Yethiraj}},
  \bibinfo{author}{\bibfnamefont{W.}~\bibnamefont{Montfrooij}},
  \bibinfo{author}{\bibfnamefont{J.~S.} \bibnamefont{Gardner}},
  \bibinfo{author}{\bibfnamefont{P.}~\bibnamefont{Metcalf}}, \bibnamefont{and}
  \bibinfo{author}{\bibfnamefont{J.~M.} \bibnamefont{Honig}},
  \bibinfo{journal}{Phys. Rev. B} \textbf{\bibinfo{volume}{73}},
  \bibinfo{pages}{174409} (\bibinfo{year}{2006}).

\bibitem[{\citenamefont{Huang et~al.}(2012)\citenamefont{Huang, Fan, Qu, Chen,
  Wang, Wu, Chen, Xiao, and Chien}}]{chien_magnproxPt}
\bibinfo{author}{\bibfnamefont{S.~Y.} \bibnamefont{Huang}},
  \bibinfo{author}{\bibfnamefont{X.}~\bibnamefont{Fan}},
  \bibinfo{author}{\bibfnamefont{D.}~\bibnamefont{Qu}},
  \bibinfo{author}{\bibfnamefont{Y.~P.} \bibnamefont{Chen}},
  \bibinfo{author}{\bibfnamefont{W.~G.} \bibnamefont{Wang}},
  \bibinfo{author}{\bibfnamefont{J.}~\bibnamefont{Wu}},
  \bibinfo{author}{\bibfnamefont{T.~Y.} \bibnamefont{Chen}},
  \bibinfo{author}{\bibfnamefont{J.~Q.} \bibnamefont{Xiao}}, \bibnamefont{and}
  \bibinfo{author}{\bibfnamefont{C.~L.} \bibnamefont{Chien}},
  \bibinfo{journal}{Phys. Rev. Lett.} \textbf{\bibinfo{volume}{109}},
  \bibinfo{pages}{107204} (\bibinfo{year}{2012}).

\bibitem[{\citenamefont{Mizuguchi et~al.}(2012)\citenamefont{Mizuguchi, Ohata,
  Uchida, Saitoh, and Takanashi}}]{ANE_FePt}
\bibinfo{author}{\bibfnamefont{M.}~\bibnamefont{Mizuguchi}},
  \bibinfo{author}{\bibfnamefont{S.}~\bibnamefont{Ohata}},
  \bibinfo{author}{\bibfnamefont{K.}~\bibnamefont{Uchida}},
  \bibinfo{author}{\bibfnamefont{E.}~\bibnamefont{Saitoh}}, \bibnamefont{and}
  \bibinfo{author}{\bibfnamefont{K.}~\bibnamefont{Takanashi}},
  \bibinfo{journal}{Appl. Phys. Express} \textbf{\bibinfo{volume}{5}},
  \bibinfo{pages}{093002} (\bibinfo{year}{2012}).

\bibitem[{\citenamefont{Lide}(2009)}]{Lide2009}
\bibinfo{author}{\bibfnamefont{D.}~\bibnamefont{Lide}},
  \emph{\bibinfo{title}{CRC Handbook of Chemistry \& Physics}}
  (\bibinfo{publisher}{Taylor \& Francis Group}, \bibinfo{year}{2009}).

\bibitem[{\citenamefont{Wilhelm et~al.}(2000)\citenamefont{Wilhelm,
  Poulopoulos, Ceballos, Wende, Baberschke, Srivastava, Benea, Ebert,
  Angelakeris, Flevaris et~al.}}]{magn_prox_NiPt}
\bibinfo{author}{\bibfnamefont{F.}~\bibnamefont{Wilhelm}},
  \bibinfo{author}{\bibfnamefont{P.}~\bibnamefont{Poulopoulos}},
  \bibinfo{author}{\bibfnamefont{G.}~\bibnamefont{Ceballos}},
  \bibinfo{author}{\bibfnamefont{H.}~\bibnamefont{Wende}},
  \bibinfo{author}{\bibfnamefont{K.}~\bibnamefont{Baberschke}},
  \bibinfo{author}{\bibfnamefont{P.}~\bibnamefont{Srivastava}},
  \bibinfo{author}{\bibfnamefont{D.}~\bibnamefont{Benea}},
  \bibinfo{author}{\bibfnamefont{H.}~\bibnamefont{Ebert}},
  \bibinfo{author}{\bibfnamefont{M.}~\bibnamefont{Angelakeris}},
  \bibinfo{author}{\bibfnamefont{N.~K.} \bibnamefont{Flevaris}},
  \bibnamefont{et~al.}, \bibinfo{journal}{Phys. Rev. Lett.}
  \textbf{\bibinfo{volume}{85}}, \bibinfo{pages}{413} (\bibinfo{year}{2000}).

\end{thebibliography}
\end{document}